\documentclass[11pt]{article}
\usepackage[margin=1in]{geometry}
\usepackage[T1]{fontenc}
\usepackage[utf8]{inputenc}
\usepackage{lmodern}
\usepackage{graphicx}
\usepackage{amsmath,amssymb}
\usepackage{booktabs,longtable,array,tabularx}
\usepackage{xcolor}
\usepackage{url}
\usepackage[colorlinks=true,linkcolor=black,citecolor=black,urlcolor=blue]{hyperref}
\hypersetup{
  pdftitle={SurvFM enables tabular foundation models for right-censored survival prediction},
  pdfauthor={Yue Lyu; Steven H. Lin; Xuelin Huang; Ziyi Li},
  pdfsubject={Right-censored survival prediction with tabular foundation models},
  pdfkeywords={survival analysis, right censoring, restricted mean survival time, tabular foundation models}
}
\usepackage{setspace}
\graphicspath{{figures/}{extended_data/}}

\DeclareUnicodeCharacter{2011}{-}
\DeclareUnicodeCharacter{2013}{--}
\DeclareUnicodeCharacter{2014}{---}
\DeclareUnicodeCharacter{2265}{$\geq$}
\DeclareUnicodeCharacter{2264}{$\leq$}
\DeclareUnicodeCharacter{2212}{-}

\title{SurvFM enables tabular foundation models for right-censored survival prediction}
\author{\parbox{0.95\textwidth}{\centering
Yue Lyu$^{1}$, Steven H. Lin$^{2}$, Xuelin Huang$^{1,*}$ and Ziyi Li$^{1,*}$\\[0.4em]
\small $^{1}$Department of Biostatistics, Division of Discovery Science, The University of Texas MD Anderson Cancer Center, Houston, Texas, USA\\
\small $^{2}$Department of Thoracic Radiation Oncology, Division of Radiation Oncology, The University of Texas MD Anderson Cancer Center, Houston, Texas, USA\\[0.4em]
\small $^{*}$Correspondence: \texttt{XLHuang@MDAnderson.org}; \texttt{ZLi16@MDAnderson.org}}}
\date{}

\begin{document}
\maketitle

\begin{abstract}
General-purpose tabular foundation models can be adapted across prediction tasks, but right censoring leaves many event times unknown and prevents their direct use as regression labels. SurvFM converts censored follow-up into observation-level targets for restricted mean survival time (RMST), the expected event-free time accumulated up to a chosen horizon. These targets allow multiple tabular foundation models to predict RMST without architectural modification. In simulations with known RMST, SurvFM achieved leading RMST accuracy and competitive discrimination across heterogeneous settings. Its targets were more accurate than simpler outcome constructions, with larger gains as censoring increased. Across 55 public datasets, SurvFM models remained in the leading performance band under full and restricted training. Models fitted in either of two public myelodysplastic syndrome cohorts retained competitive performance in the other without refitting. SurvFM separates censoring handling from prediction architecture, allowing advances in general-purpose tabular prediction to enter survival analysis without model-specific redesign.
\end{abstract}

\section{Introduction}

Foundation models offer general-purpose prediction machinery that can be adapted across datasets and tasks, reducing the need to design a new model for every application. In structured data, tabular foundation models apply this principle to classification and regression with limited task-specific adaptation \cite{hollmann2025tabpfn}. Ordinary regression nevertheless requires a usable outcome label for every training record. Right-censored time-to-event data violate this requirement: for a censored observation, the event time is unobserved and known only to exceed the observed follow-up. This creates a direct mismatch between the information in a censored outcome and the labels required by supervised regression.

Established survival methods resolve this mismatch by incorporating censoring within the estimator, objective or architecture. Cox regression, random survival forests, survival ensembles and neural survival models exemplify this strategy \cite{cox1972regression,ishwaran2008random,hothorn2006survivalensembles,katzman2018deepsurv,kvamme2019pycox,lyu2026supersurv}. Recent studies have similarly explored survival-specific adaptations of tabular foundation models \cite{kim2026tabularsurvival,qi2026survivalpfn,boehm2026survpfn}. These approaches couple censoring handling to a particular modelling procedure. Each new tabular model may therefore require its own objective, output representation or fitting strategy, making it difficult to separate the contribution of the prediction architecture from that of the censored-outcome construction.

An alternative is to adapt the outcome representation while leaving the prediction engine unchanged. Covariate-conditional restricted mean survival time (RMST) provides a natural prognostic target: the expected event-free time accumulated up to a prespecified horizon among observations with similar baseline characteristics \cite{royston2013rmst,uno2014beyondhazardratio}. It is a bounded conditional mean expressed in the original time units and therefore defines a standard regression estimand, but censoring prevents its direct observation for each training record. Jackknife pseudo-observations use the full censored sample and leave-one-out estimates to construct observation-level quantities whose conditional mean targets RMST under the required censoring conditions \cite{andersen2004rmstpseudo,andersen2010pseudoobservations,parner2010pseudoobservations,cwiling2025superelearner}. This construction suggests an estimand-aligned route from censored outcomes to general-purpose regression, but its ability to serve as a common representation across modern tabular models has not been systematically established.

Here we introduce SurvFM (\textbf{Surv}ival \textbf{F}oundation \textbf{M}odels), a computational framework that pairs pseudo-RMST outcome representations with multiple tabular foundation models. SurvFM places the survival-specific operation in target construction and poses the remaining task as horizon-specific regression, without requiring a survival-specific architecture or loss function for each tabular model. We test the representation in paired censoring interventions and simulations with known RMST, evaluate five SurvFM models within a 17-model benchmark spanning 55 eligible static SurvSet datasets \cite{drysdale2022survset}, and examine direct bidirectional transport between two public myelodysplastic syndrome cohorts with harmonized clinical and mutation features \cite{bernard2022ipssm,reilly2019mdsmethylation}. This design separates the effects of target construction and prediction architecture while testing transfer across tabular models, data regimes and populations.

\section{Results}

\subsection{SurvFM converts censored follow-up into regression targets for tabular models}

\begin{figure}[p]
\centering
\includegraphics[width=\linewidth,height=0.50\textheight,keepaspectratio]{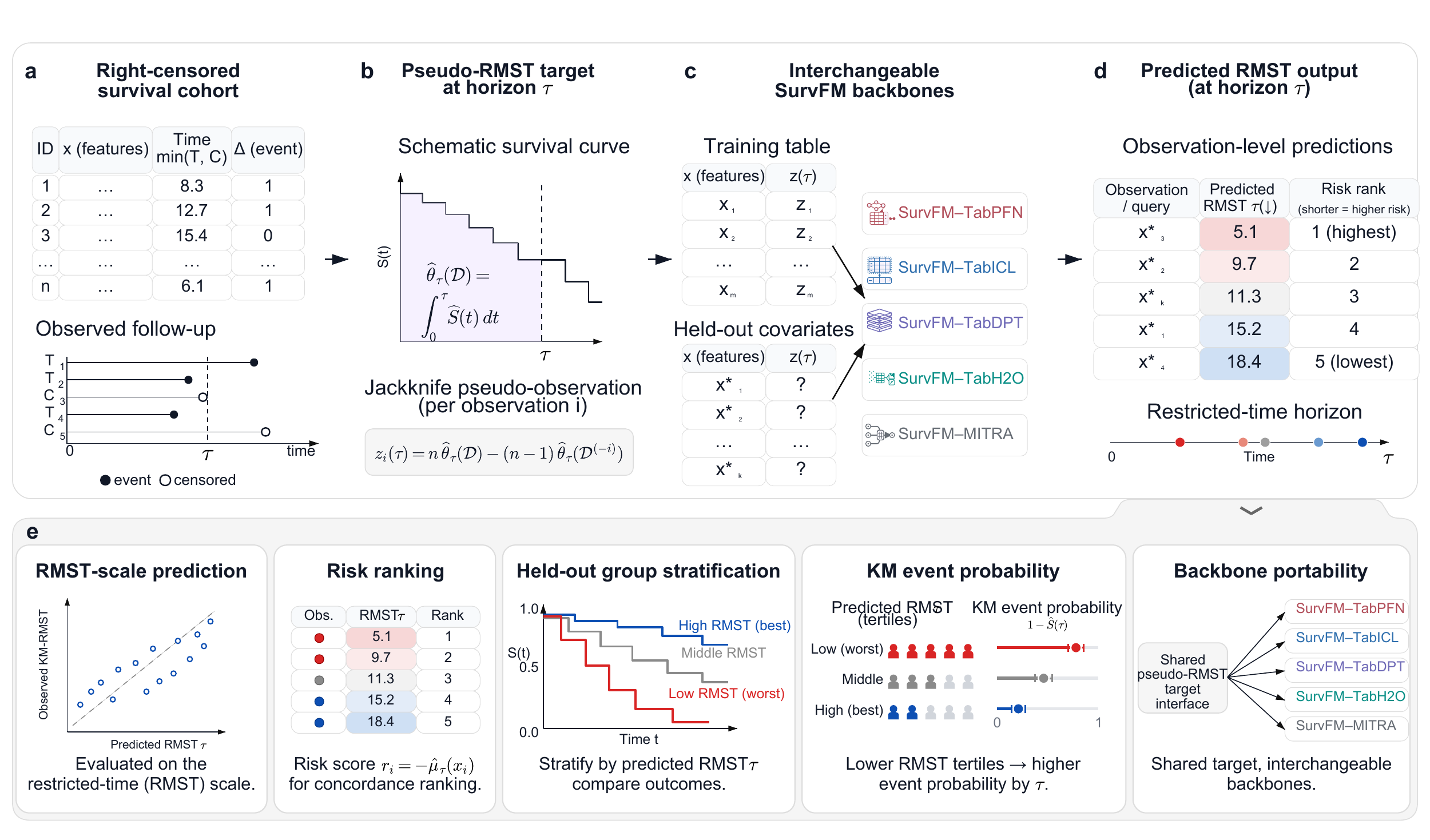}
\caption{\textbf{SurvFM converts censored follow-up into RMST targets for tabular foundation models.} \textbf{a,} Right-censored tabular data contain baseline features, observed follow-up $Y_i=\min(T_i,C_i)$ and event indicator $\Delta_i$; the event time remains unknown when $\Delta_i=0$. \textbf{b,} At horizon $\tau$, $\widehat\theta_\tau(\mathcal D)$ is the marginal Kaplan--Meier RMST functional. Jackknife pseudo-observations $z_i(\tau)=n\widehat\theta_\tau(\mathcal D)-(n-1)\widehat\theta_\tau(\mathcal D_{-i})$ provide one target for expected event-free time per training observation while retaining censored records. \textbf{c,} The same pseudo-RMST targets are supplied to five tabular foundation models. \textbf{d,} Each model produces an observation-level RMST prediction at $\tau$; shorter predicted RMST corresponds to higher risk for discrimination. \textbf{e,} Readouts include RMST-scale error, risk ranking, retrospective group-level outcome stratification, Kaplan--Meier event probability and portability across tabular models. Observed Kaplan--Meier RMST and event probability are group-level quantities, not directly observed individual outcomes. Numerical values are illustrative.}
\label{fig:figure1}
\end{figure}

SurvFM resolves the label mismatch before fitting the prediction model (\textbf{Fig.~\ref{fig:figure1}}). From baseline covariates, observed follow-up and event indicators, it constructs jackknife pseudo-observations for RMST at a prespecified prediction horizon, denoted $\tau$. This produces one target for expected event-free time per training record while retaining censored observations without treating their follow-up as a completed event time. The selected tabular model, hereafter the backbone, then learns to predict RMST from covariates (Methods, Section~\ref{sec:methods-pseudo-rmst}).

The same pseudo-RMST targets were supplied to SurvFM-TabPFN, SurvFM-TabICL, SurvFM-TabDPT, SurvFM-TabH2O and SurvFM-MITRA; SurvFM-TabPFN denotes TabPFN-v3 unless stated otherwise. Four conventional regressors trained on the same targets and eight survival-model comparators formed a 17-model benchmark. Each SurvFM backbone outputs horizon-specific restricted event-free time, rather than a full survival curve. Negative predicted RMST was used as a risk score for discrimination, so shorter predicted RMST corresponded to higher predicted risk (Methods, Eq.~\ref{eq:risk-orientation}).

\subsection{SurvFM performs competitively across heterogeneous simulation settings}

\begin{figure}[p]
\centering
\includegraphics[width=\linewidth,height=0.68\textheight,keepaspectratio]{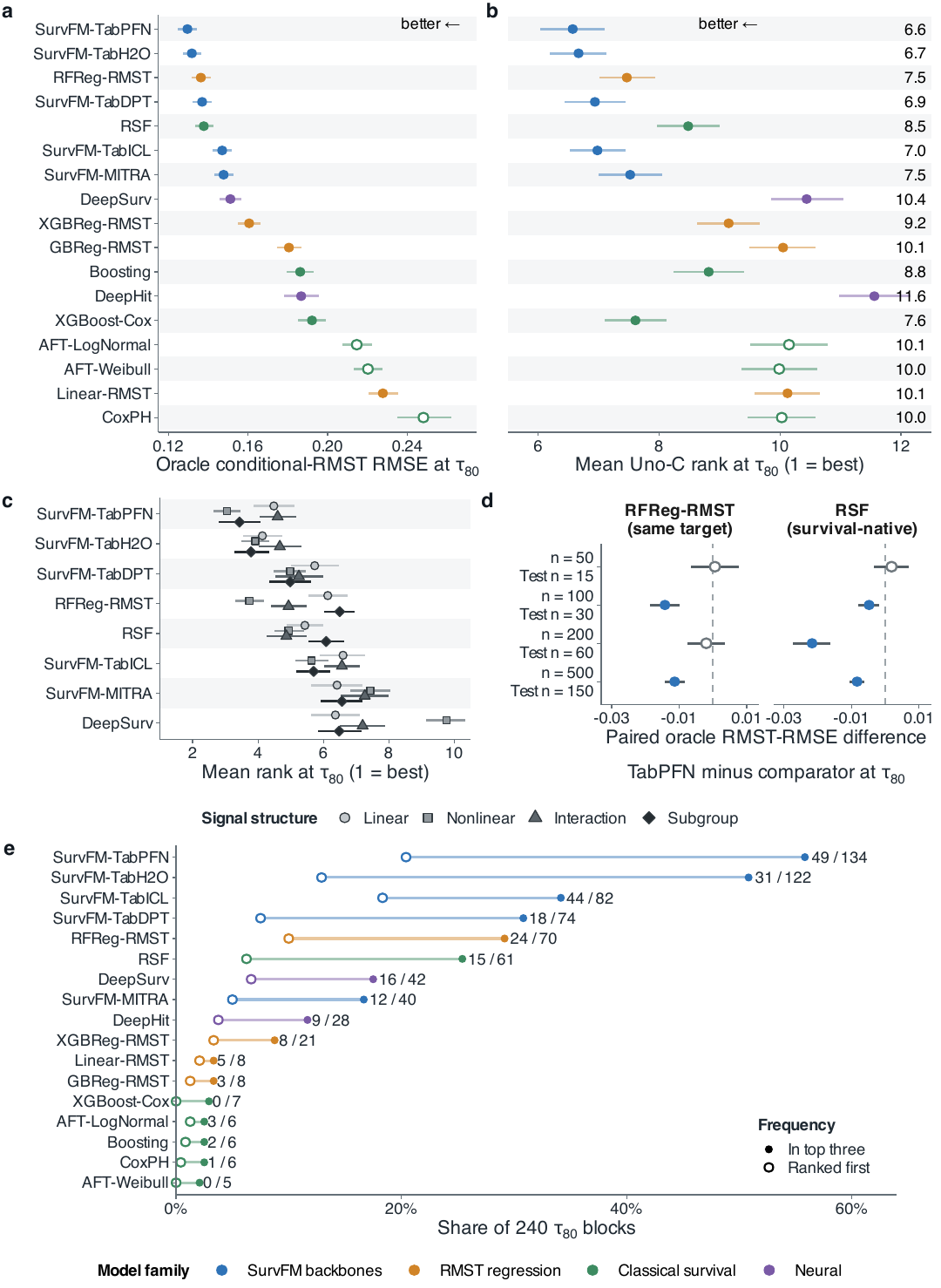}
\caption{\textbf{SurvFM performance across simulation settings with known RMST.} The benchmark comprised 24 settings, ten replicates, three prediction horizons and 17 models. Ranks were computed within each setting-replicate block among evaluable models; failed or non-evaluable rows were not imputed. \textbf{a,} Mean oracle conditional-RMST RMSE at $\tau_{80}$. Points are means over evaluable blocks and bars are 95\% setting-cluster bootstrap intervals. Open markers identify models with reduced fit coverage. \textbf{b,} Mean within-block Uno's C-index rank at $\tau_{80}$, retaining the panel-a model order. Numbers give point-estimate mean ranks. \textbf{c,} Signal-structure-specific oracle-error ranks for five SurvFM models and the three highest-ranked comparators by overall mean rank. \textbf{d,} Paired oracle RMST-RMSE difference between SurvFM-TabPFN and RFReg-RMST or random survival forest across sample sizes. Negative values favour SurvFM-TabPFN; filled markers denote intervals excluding zero. \textbf{e,} Shares of 240 $\tau_{80}$ blocks in which each model ranked first or in the top three; labels give first-place and top-three counts. Models are ordered by top-three frequency, with first-place frequency and overall mean rank used as successive tie-breakers.}
\label{fig:figure2}
\end{figure}

Because true conditional RMST is unknown in real data, we also evaluated SurvFM in 24 simulation settings with a known conditional survival function (\textbf{Fig.~\ref{fig:figure2}}, \textbf{Extended Data Fig.~1} and \textbf{Supplementary Fig.~2}). Numerical integration of that function provided the true RMST for each baseline profile. The design used ten replicates per setting and feature counts of $p=10$, $50$ and $99$. We calculated the root-mean-square error (RMSE) between this known truth and held-out predictions. We refer to this simulation-only metric as oracle conditional-RMST RMSE; oracle describes the evaluation target, not information supplied to the fitted model (Methods, Eq.~\ref{eq:rmse-oracle}).

At $\tau_{80}$, SurvFM-TabPFN had the lowest mean oracle RMSE among evaluable blocks (0.1296), followed by SurvFM-TabH2O (0.1318), random-forest RMST regression (RFReg-RMST; 0.1364), SurvFM-TabDPT (0.1370) and random survival forest (RSF; 0.1378; \textbf{Fig.~\ref{fig:figure2}a}). Mean ranks by Uno's C-index (Uno-C) placed SurvFM-TabPFN, SurvFM-TabH2O, SurvFM-TabDPT and SurvFM-TabICL in a leading group, with means from 6.57 to 6.98 among evaluable models (\textbf{Fig.~\ref{fig:figure2}b}). Accuracy and discrimination produced different model orderings, but the SurvFM family remained concentrated toward the top under both criteria. RMST error evaluates numerical recovery on the restricted-time scale, whereas Uno-C evaluates risk ordering.

The regime analysis used 240 $\tau_{80}$ setting-replicate blocks to examine performance across signal structures and sample sizes (\textbf{Fig.~\ref{fig:figure2}c--e}; Methods, Section~\ref{sec:methods-simulation-summaries}). Panel c displays the five SurvFM backbones and the three highest-ranked comparators by overall mean rank; ranks were calculated against all 17 models, whose complete regime summaries are shown in \textbf{Supplementary Fig.~1}. SurvFM-TabPFN had the best overall mean oracle-error rank (3.89) and remained within a mean rank of 4.60 in its least-favourable signal structure. SurvFM-TabH2O followed with overall and least-favourable ranks of 4.13 and 4.67. RFReg-RMST and RSF remained competitive in selected small-sample or signal settings, indicating that comparator performance was regime-dependent (\textbf{Fig.~\ref{fig:figure2}c} and \textbf{Supplementary Fig.~2b}).

Direct comparisons across sample sizes evaluated SurvFM-TabPFN against RFReg-RMST and RSF (\textbf{Fig.~\ref{fig:figure2}d}). At $n=50$, its oracle RMST error was comparable with both methods. At $n=100$, it had lower error than both; at $n=200$, it had lower error than RSF and remained comparable with RFReg-RMST. It also had lower error than both comparators at $n=500$. Across the smaller-sample settings ($n=50$--200), SurvFM-TabPFN was therefore comparable with or better than both key comparators. Across all blocks, SurvFM-TabPFN ranked first in 49 of 240 blocks and in the top three in 134; SurvFM-TabH2O did so in 31 and 122, respectively (\textbf{Fig.~\ref{fig:figure2}e}). SurvFM-TabPFN and SurvFM-TabH2O thus combined strong ranks across signal structures with frequent top-three placement across heterogeneous controlled regimes.

\subsection{Pseudo-RMST targets improve recovery as censoring increases}

\begin{figure}[p]
\centering
\includegraphics[width=\linewidth,height=0.55\textheight,keepaspectratio]{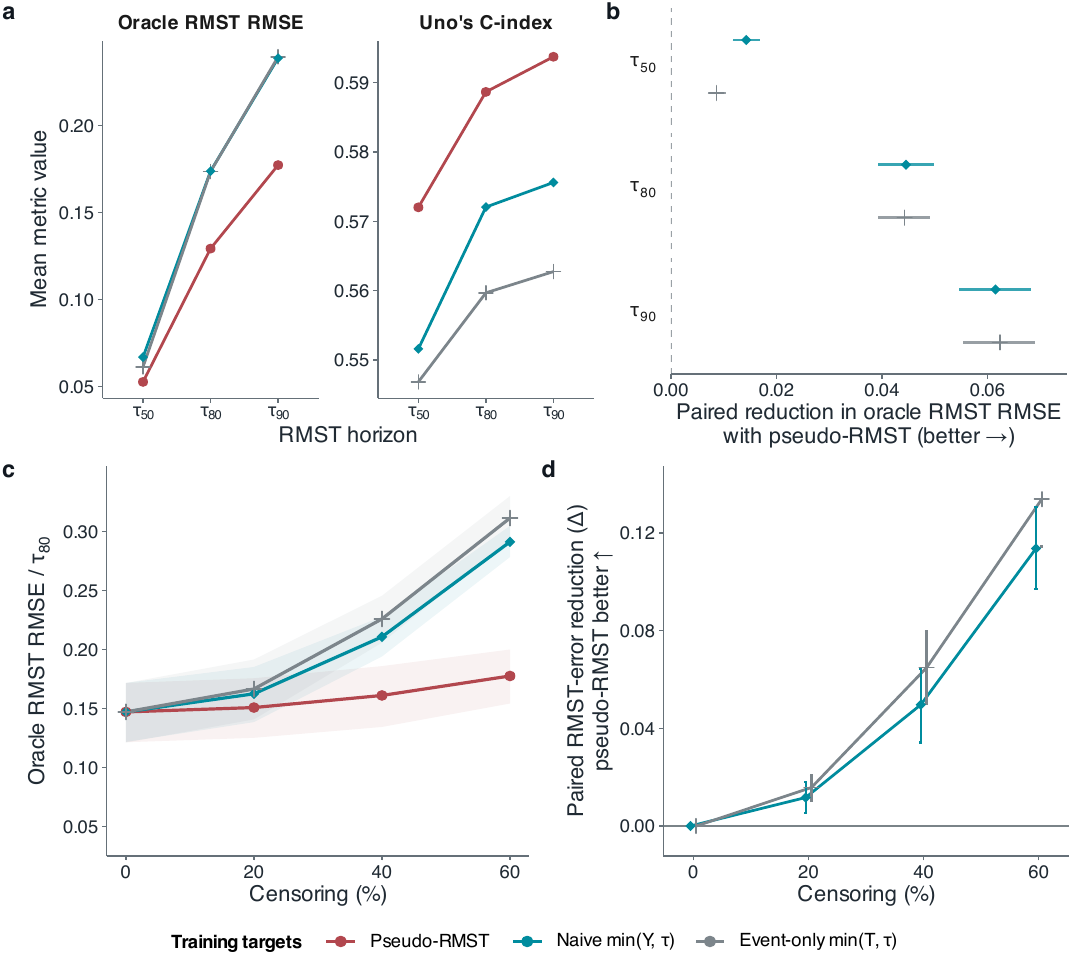}
\caption{\textbf{Pseudo-RMST target construction improves restricted-time recovery as censoring increases.} \textbf{a,} Mean oracle RMST RMSE and Uno's C-index for SurvFM-TabPFN trained with projected pseudo-RMST, naive restricted observed time or event-only restricted time at $\tau_{50}$, $\tau_{80}$ and $\tau_{90}$ in the 24-setting target-ablation experiment. \textbf{b,} Paired reduction in oracle RMST RMSE for pseudo-RMST relative to the two alternative targets; positive values favour pseudo-RMST. Points are mean paired differences and bars are 95\% setting-cluster bootstrap intervals. \textbf{c,} Mean oracle RMST RMSE divided by $\tau_{80}$ at exactly 0\%, 20\%, 40\% and 60\% censoring in paired simulations holding covariates, latent event times, splits and horizons fixed. Lines show projected pseudo-RMST, naive $\min(Y,\tau)$ and event-only $\min(T,\tau)$ targets; ribbons are 95\% setting-cluster bootstrap intervals. \textbf{d,} Paired RMST-error reduction for pseudo-RMST relative to the two alternatives in the censoring intervention. The complete-data and conditional-RMST reference arms and latent-time concordance sensitivity are shown in Extended Data Fig.~2.}
\label{fig:figure3}
\end{figure}

With the SurvFM-TabPFN backbone held fixed in a separate ten-replicate ablation, pseudo-RMST yielded lower oracle RMSE and higher Uno-C than naive restricted observed time and event-only targets at $\tau_{50}$, $\tau_{80}$ and $\tau_{90}$ (\textbf{Fig.~\ref{fig:figure3}a,b}). All 12 paired 95\% bootstrap intervals obtained by resampling the 24 settings excluded zero. Naive observed-time labels treat censoring time as completed follow-up, whereas event-only training excludes censored observations. Jackknife pseudo-observations instead retain censored observations in the RMST regression task. A bounded-versus-raw target sensitivity is reported in \textbf{Supplementary Table~9}.

We next varied censoring while holding the latent prediction problem fixed. Twelve simulation settings and ten replicates were evaluated at exactly 0\%, 20\%, 40\% and 60\% censoring using shared covariates, latent event times, train--test assignments and $\tau_{80}$ horizons. SurvFM-TabPFN and SurvFM-TabICL were each trained with projected pseudo-RMST, naive restricted observed time, event-only restricted time, complete-data restricted event time and the known conditional-RMST target. The latter two arms are nondeployable references shown in Extended Data Fig.~2a: complete-data training reveals every event time, whereas the conditional-RMST arm supplies the estimand itself as the training target (Methods, Section~\ref{sec:methods-paired-censoring}).

Without censoring, pseudo-RMST, naive observed time and complete-data restricted time were identical by construction. As censoring increased, pseudo-RMST recovered progressively more of the information lost by naive outcome construction (\textbf{Fig.~\ref{fig:figure3}c,d}). At 60\% censoring, its mean reduction in tau-normalized oracle RMST RMSE relative to naive restricted observed time was 0.1136 (95\% confidence interval (CI), 0.0970--0.1307) for SurvFM-TabPFN and 0.1049 (0.0802--0.1288) for SurvFM-TabICL. Pseudo-RMST also had lower error than the event-only target at every nonzero censoring level for both backbones (\textbf{Fig.~\ref{fig:figure3}d}).

At 60\% censoring, pseudo-RMST avoided 79\% and 76\% of the naive target's excess error over complete-data training for SurvFM-TabPFN and SurvFM-TabICL, respectively, while remaining above the complete-data and conditional-RMST reference arms (\textbf{Extended Data Fig.~2a}). Latent-time concordance was unchanged without censoring and generally improved relative to naive training as censoring increased (\textbf{Extended Data Fig.~2b}). Uno-C showed the same broad pattern where evaluable; dynamic AUC was too sparsely supported at 60\% censoring for a primary conclusion. The paired intervention therefore attributes an increasing fraction of the observed advantage to censoring-aware outcome construction rather than to a change in prediction architecture.

\subsection{SurvFM models rank highly across the 55-dataset SurvSet benchmark}

\begin{figure}[p]
\centering
\includegraphics[width=\linewidth,height=0.51\textheight,keepaspectratio]{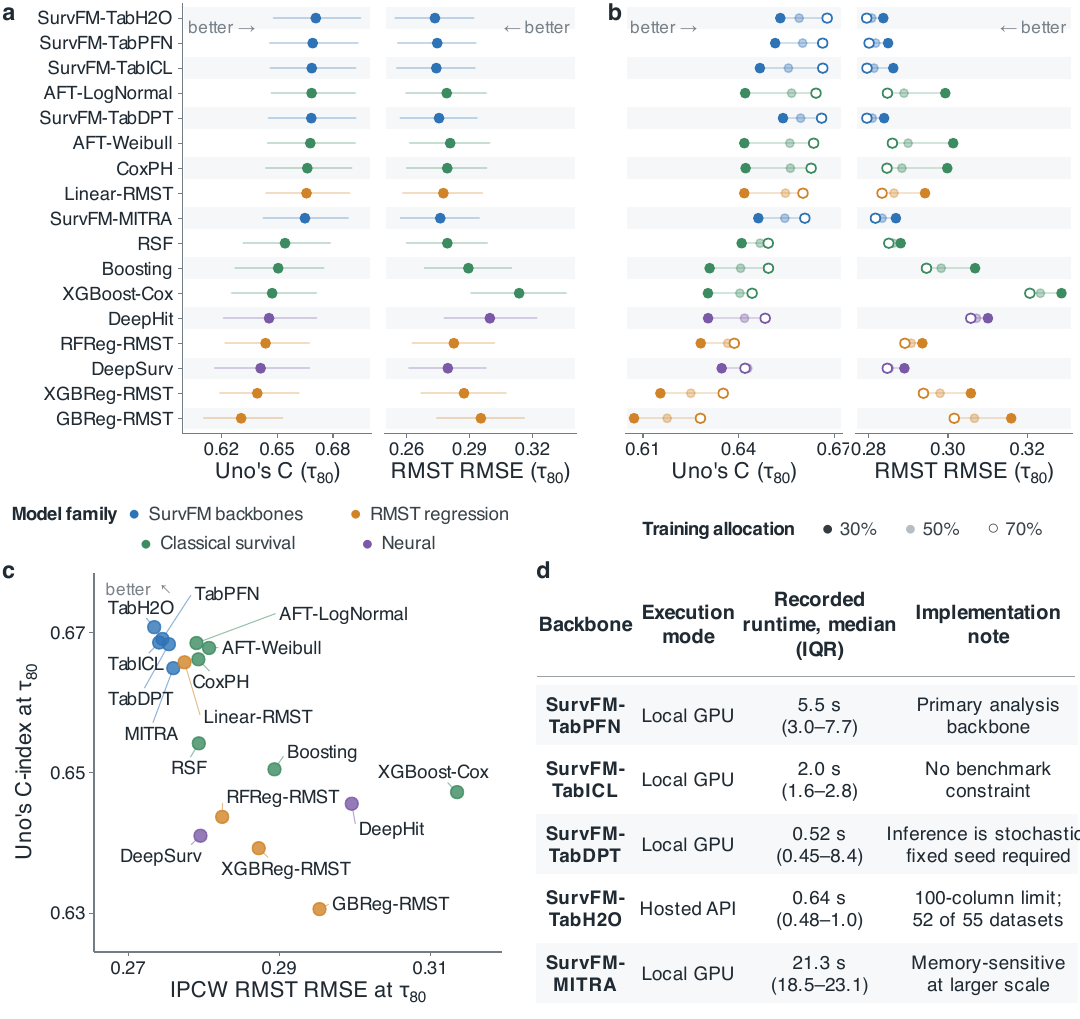}
\caption{\textbf{Public-data performance, training allocation and SurvFM model profiles.} The benchmark comprised 55 eligible static SurvSet datasets, ten fixed splits and 17 models. \textbf{a,} Performance across the 51 datasets for which all models were evaluable, with models ordered by decreasing mean Uno-C at the training-derived $\tau_{80}$ horizon. Points are dataset-weighted means; bars are 95\% percentile bootstrap intervals obtained by resampling datasets while retaining all ten splits together. Uno-C and tau-normalized IPCW RMST RMSE were evaluable in 476 and 497 blocks. \textbf{b,} Nested training-allocation sensitivity, retaining the panel-a model order. Subsets containing 30\% and 50\% of the full cohort were nested within each 70\% outer training pool while the test fold, horizon and censoring reference remained fixed. Summaries using only blocks evaluable for all models contained 426 Uno-C and 447 RMST-error blocks from 47 datasets. \textbf{c,} Joint $\tau_{80}$ performance using panel-a means; higher Uno-C and lower RMST error are favourable. \textbf{d,} Descriptive profiles of the five SurvFM models. Runtime is median recorded fit and prediction time per successful row, not a controlled speed comparison. TabH2O completed 52 of 55 datasets because three exceeded its 100-processed-column limit. Missing combinations were not imputed.}
\label{fig:figure4}
\end{figure}

The public-data benchmark comprised 55 eligible static SurvSet datasets, ten fixed 70:30 splits per dataset, three horizons and 17 models (\textbf{Fig.~\ref{fig:figure4}}). Primary $\tau_{80}$ comparisons used the 51 datasets on which all models could be evaluated, fixing the comparison set while retaining most datasets.

Four SurvFM backbones occupied the leading discrimination band. Mean Uno-C was 0.6707 for SurvFM-TabH2O, 0.6691 for SurvFM-TabPFN, 0.6686 for SurvFM-TabICL and 0.6683 for SurvFM-TabDPT, interleaved with the log-normal accelerated failure-time model (AFT-LogNormal; 0.6685) and other strong classical models (\textbf{Fig.~\ref{fig:figure4}a}). SurvFM-MITRA ranked ninth at 0.6649. To compare RMST-scale accuracy across datasets with different follow-up scales, we used inverse-probability-of-censoring-weighted (IPCW) RMST RMSE normalized by $\tau$ (Methods, Eq.~\ref{eq:ipcw-rmst-rmse}). Under this lower-is-better metric, SurvFM-TabH2O, SurvFM-TabICL, SurvFM-TabPFN, SurvFM-TabDPT and SurvFM-MITRA occupied the five lowest point estimates (0.2734--0.2760; \textbf{Fig.~\ref{fig:figure4}a,c}). Although the model-specific bootstrap intervals overlapped, paired within-dataset contrasts were more precise because every model was evaluated on the same datasets (\textbf{Supplementary Table~6}). Rankings at $\tau_{90}$ were highly similar, whereas at $\tau_{50}$ several classical models achieved higher discrimination (\textbf{Supplementary Fig.~3b} and \textbf{Supplementary Table~4}); model ordering therefore depended partly on the prediction horizon.

SurvFM backbones retained the strongest performance band when training was restricted to 30\% of the full cohort (\textbf{Fig.~\ref{fig:figure4}b}; Methods, Section~\ref{sec:methods-training-allocation}). Across the 47 datasets for which all 17 models were evaluable, the five SurvFM backbones had the five highest mean Uno-C values (0.6461--0.6539) and the five lowest normalized RMST errors (0.2837--0.2868) at 30\% allocation. The family remained concentrated near the top at 50\% and 70\% allocation. The same outer test fold, horizon and censoring weights were retained across allocations, with the 30\% and 50\% training subsets nested inside the full 70\% training set.

The five backbones also differed in how they could be deployed (\textbf{Fig.~\ref{fig:figure4}d}). Because SurvFM-TabPFN combined consistently strong performance, local execution and complete dataset coverage, we used it for the detailed public-data analyses. SurvFM-TabICL and SurvFM-TabDPT provided additional local alternatives with complete coverage. SurvFM-TabH2O frequently achieved leading performance, but required a hosted application programming interface (API) and was limited to 52 of 55 datasets by its 100-processed-column constraint. SurvFM-MITRA completed all datasets but was more computationally and memory intensive in our implementation. The shared pseudo-RMST construction therefore supports backbone selection according to predictive performance, dimensionality, execution environment and available resources.

\subsection{Predicted RMST expresses prognosis in event-free-time units and stratifies observed outcomes}

\begin{figure}[p]
\centering
\includegraphics[width=\linewidth,height=0.50\textheight,keepaspectratio]{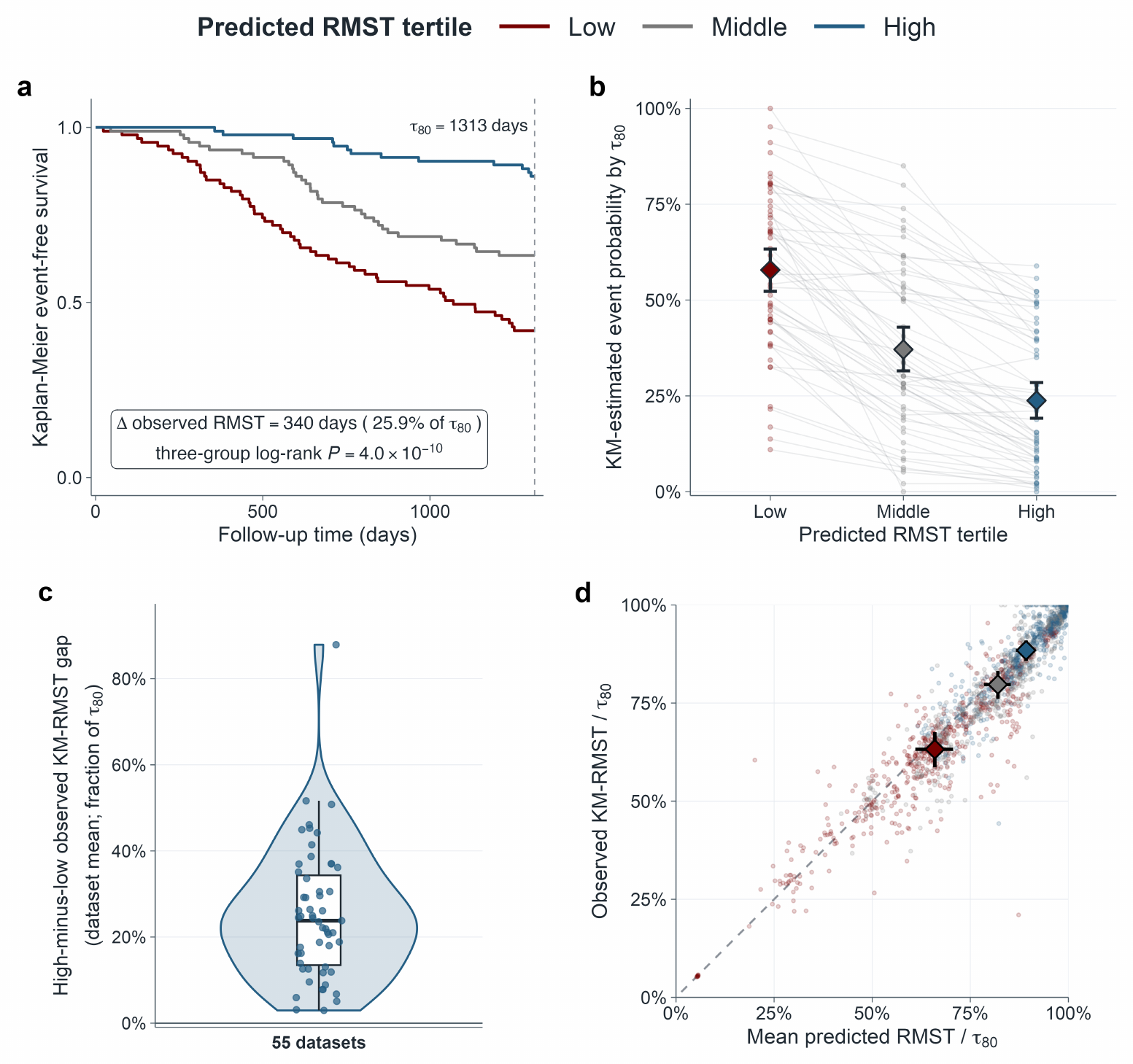}
\caption{\textbf{Predicted RMST stratifies held-out event-free time and event probability.} SurvFM-TabPFN v3 produced predictions targeting covariate-conditional RMST at the training-derived $\tau_{80}$ horizon across 55 datasets and ten splits. Held-out observations were divided into low, middle and high predicted-RMST tertiles. \textbf{a,} Kaplan--Meier curves for Colon split 1, retained from a fixed illustrative set; its normalized gap ranked 32nd of 55. The annotation reports the high-minus-low observed KM-RMST gap and descriptive omnibus log-rank result. \textbf{b,} KM-estimated event probability by $\tau_{80}$. Thin lines are dataset trajectories after averaging over splits; diamonds and bars show cross-dataset means and 95\% percentile bootstrap intervals obtained by resampling datasets. \textbf{c,} Dataset-mean high-minus-low observed KM-RMST gaps. All 55 means were positive, while 537 of 550 split-level gaps were positive; observed KM-RMST was strictly increasing across tertiles in 446 blocks and non-decreasing in 462. \textbf{d,} Grouped predicted-versus-observed RMST agreement. Small points are 1,650 non-independent dataset, split and tertile summaries; diamonds and orthogonal bars summarize dataset-averaged tertile means and 95\% percentile bootstrap intervals obtained by resampling datasets. Observed RMST was below prediction by 2.74, 2.29 and 0.77 percentage points of $\tau_{80}$ in the low, middle and high tertiles. Thirty-six groups had no observations at risk at $\tau_{80}$; survival had reached zero in 27, whereas 9 groups from 7 blocks retained unsupported positive tails. Ordering and minimum-at-risk diagnostics are shown in \textbf{Supplementary Fig.~6}. These retrospective analyses do not establish individual calibration or clinical utility.}
\label{fig:figure5}
\end{figure}

RMST summarizes expected event-free time up to a chosen horizon and, unlike event probability at the horizon, reflects when events occur within that interval. We therefore evaluated whether SurvFM predictions preserved prognostic information in test data through group-level RMST agreement and outcome stratification (\textbf{Fig.~\ref{fig:figure5}}; Methods, Section~\ref{sec:methods-rmst-stratification}). Within each dataset and split, test records not used for model fitting were assigned to low, middle and high predicted-RMST tertiles, and their observed event-free survival and event probability were compared. The tertiles evaluated outcome separation; the underlying predictions remained continuous. Colon provided a fixed illustrative example near the median cross-dataset separation, while the primary summaries aggregated all 55 datasets and ten splits.

Kaplan--Meier (KM)-estimated event probability by $\tau_{80}$ was lower in the high than the low predicted-RMST tertile in all 55 datasets after averaging over splits, and was non-increasing from low to middle to high in 50 datasets, allowing ties (\textbf{Fig.~\ref{fig:figure5}b}). Dataset-mean high-minus-low observed KM-RMST gaps were positive in all 55 datasets, with a median gap of 23.8\% of $\tau_{80}$ (\textbf{Fig.~\ref{fig:figure5}c}). At the split level, 537 of 550 gaps were positive; observed KM-RMST was strictly increasing across tertiles in 446 blocks and non-decreasing in 462.

Across tertiles, the median block-level mean absolute difference between predicted and Kaplan--Meier-estimated RMST was 3.3\% of $\tau_{80}$ (\textbf{Fig.~\ref{fig:figure5}d}). Dataset-averaged observed RMST was below prediction by 2.74, 2.29 and 0.77 percentage points of $\tau_{80}$ in the low, middle and high tertiles, respectively, indicating modest average over-prediction. Follow-up support was incomplete in a small subset of groups and is reported with ordering and minimum-at-risk diagnostics in \textbf{Supplementary Fig.~6}. These analyses establish retrospective group-level stratification and agreement. Individual calibration, treatment selection and deployment utility were not evaluated.

\subsection{SurvFM retains competitive prognostic performance across public MDS cohorts}

\begin{figure}[p]
\centering
\includegraphics[width=\linewidth,height=0.54\textheight,keepaspectratio]{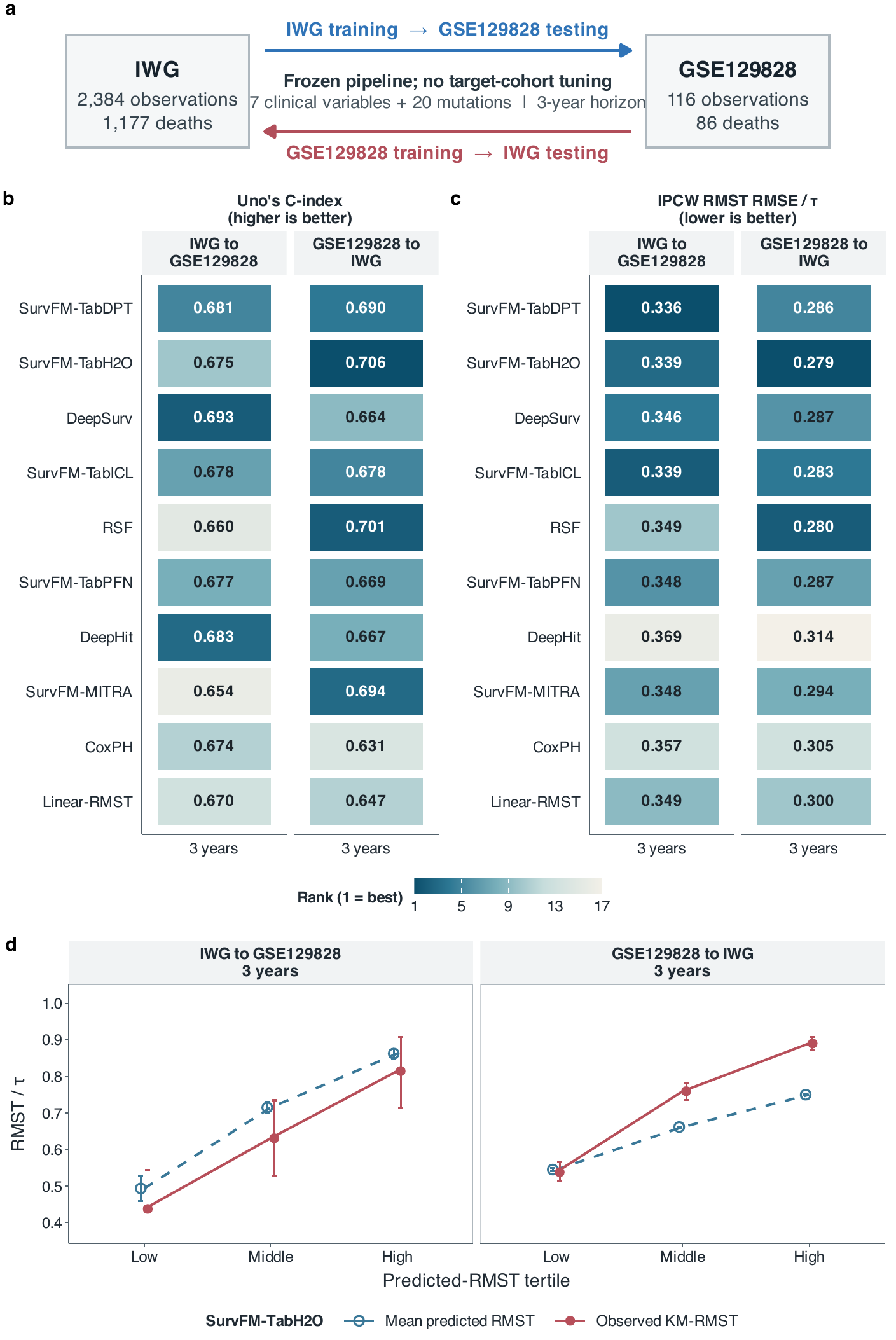}
\caption{\textbf{SurvFM transports between public myelodysplastic syndrome cohorts without refitting.} Models were fitted in one cohort and applied without target-cohort refitting to the other cohort using seven harmonized clinical variables, 20 mutation indicators and a fixed three-year overall-survival horizon. \textbf{a,} Bidirectional cross-cohort design. The International Working Group for Prognosis in MDS (IWG) cohort included 2,384 observations and 1,177 deaths; GSE129828 included 116 observations and 86 deaths. Preprocessing, pseudo-RMST target construction and model fitting were estimated in the source cohort and retained unchanged during transport. \textbf{b,c,} Three-year Uno's C-index and inverse-probability-of-censoring-weighted RMST RMSE divided by $\tau$ in the molecularly augmented analysis. The ten displayed models had the lowest mean joint rank across both feature sets, directions, horizons and metrics; complete 17-model results are shown in Supplementary Fig.~7. Cells report target-cohort point estimates and colour denotes within-block rank among all 17 models. Higher Uno-C and lower RMST RMSE are favourable. A SurvFM backbone led three of four displayed cells. No leader-versus-runner paired bootstrap interval excluded zero, indicating a competitive leading band rather than a single dominant model. \textbf{d,} Mean predicted and observed Kaplan--Meier RMST at three years, divided by the horizon, across predicted-RMST tertiles for SurvFM-TabH2O, which had the lowest mean molecular-transport rank among the five SurvFM backbones. Observed RMST was strictly ordered in both directions. Bars are target-observation bootstrap 95\% percentile intervals with the transported source model held fixed and exclude source-refitting uncertainty. This panel assesses grouped prognostic ordering and agreement, not individual calibration.}
\label{fig:figure6}
\end{figure}

Cross-cohort prediction asks whether a fitted model remains informative in a population whose outcomes were not used for model development. We tested this question in two separately assembled public MDS cohorts with harmonized clinical and mutation variables (\textbf{Fig.~\ref{fig:figure6}}). MDS prognostication increasingly incorporates molecular scoring systems, including IPSS-M and its evaluation in treatment-defined populations \cite{bernard2022ipssm,chien2026molecularscoring}. The International Working Group for Prognosis in MDS (IWG) cohort contributed 2,384 eligible observations and 1,177 deaths; GSE129828 contributed 116 observations and 86 deaths. Preprocessing, pseudo-RMST target construction and model fitting used only the source cohort before application to the other cohort without refitting or outcome-guided adaptation. Both directions used post-sampling overall survival at fixed two- and three-year horizons (Methods, Section~\ref{sec:methods-mds-transport}).

All 17 models were compared using seven clinical variables alone and with 20 shared mutation indicators. Across the 16 direction, feature-set, horizon and primary-metric combinations, a SurvFM backbone had the leading point estimate in seven. The main display shows the molecularly augmented three-year analysis (\textbf{Fig.~\ref{fig:figure6}b,c}). SurvFM-TabDPT had the lowest normalized RMST error for IWG-to-GSE129828 transport (0.336). In the reverse direction, SurvFM-TabH2O had the highest Uno-C (0.706) and lowest normalized RMST error (0.279); XGBoost-Cox had the highest IWG-to-GSE129828 Uno-C (0.696). A SurvFM backbone therefore led three of four displayed cells. No leader-versus-runner paired bootstrap interval excluded zero, placing these models in a competitive leading band rather than identifying a single dominant method. The complete 17-model rankings are in \textbf{Supplementary Fig.~7}.

For SurvFM-TabH2O, the backbone with the lowest mean molecular-transport rank, observed Kaplan--Meier RMST increased strictly across predicted-RMST tertiles in both directions at three years (\textbf{Fig.~\ref{fig:figure6}d}). Strict ordering occurred in 131 of 136 analyses across all models, directions, feature sets and horizons. Unlike repeated held-out validation, this analysis used no target-cohort outcomes for preprocessing, target construction, model fitting or tuning. SurvFM retained competitive discrimination, RMST-scale performance and prognostic ordering under direct cross-cohort transport. Because cohort identity and source-sample size changed together, the bidirectional comparison does not isolate their effects.

\section{Discussion}

SurvFM addresses a basic incompatibility: tabular regression requires observed outcome labels, whereas right censoring leaves many event times unknown. It resolves this mismatch before model fitting by constructing estimand-aligned RMST targets and presenting the remaining problem as horizon-specific tabular regression. Multiple general-purpose tabular models can therefore learn from the same censored outcomes without requiring a different survival objective or architecture for each model. This separation of target construction from prediction architecture is the central contribution of SurvFM. It also yields covariate-conditional RMST, expressed as expected event-free time accumulated up to a chosen horizon.

Simulations provided complementary evidence for performance and mechanism. Across heterogeneous settings, SurvFM models remained in the leading performance bands for RMST recovery and discrimination, including with limited training data. When the prediction architecture and underlying prediction problem were held fixed, pseudo-RMST outperformed simpler target constructions across horizons, with an increasing advantage as censoring increased. Together, these findings indicate that SurvFM benefits both from capable tabular models and from information recovered through censoring-aware target construction.

This evidence extended beyond simulation. Multiple SurvFM models remained competitive across 55 public datasets and restricted training allocations. The MDS analysis posed a more stringent question than another random split by applying each source-fitted model to a different cohort without using target-cohort outcomes for model development. Although no model was uniformly best, consistency across tabular models, data regimes and populations supports SurvFM as a general interface rather than a method tied to one predictor. The open implementation maintains this separation among target, horizon and tabular model.

RMST adds scientific meaning to this common learning problem. Because it accumulates event-free time over the full interval to the horizon, it distinguishes profiles with similar endpoint event probability but different event timing. Predictions express expected event-free months or years for a baseline profile; they are conditional means rather than individual event dates. Across held-out datasets, higher predicted RMST identified groups with longer observed event-free survival and lower event probability; the same ordering was largely retained in the transported MDS analyses. These results establish time-scaled prognostic stratification and group-level agreement. Individual calibration and decision utility require evaluation in the intended application.

Two considerations define the present computational scope. Each prediction applies to a chosen horizon, which should match the scientific question and have adequate follow-up support. The marginal pseudo-RMST construction also requires censoring assumptions. A bounded diagnostic found no multiplicity-adjusted evidence of systematic target degradation under the examined covariate-dependent censoring mechanisms (\textbf{Supplementary Table~8}), but does not establish validity under arbitrary informative censoring. When loss to follow-up depends more broadly on prognosis or model covariates, a conditional or inverse-probability-weighted target may be needed. Coherent multi-horizon targets, censoring-adjusted pseudo-observations and prediction uncertainty are natural extensions. More broadly, SurvFM shows that censored prediction can be adapted at the level of the outcome representation, allowing advances in general-purpose tabular learning to enter survival analysis without redesigning each model around censoring.

\section{Methods}

\subsection{Study design, data sources and prediction estimand}
\label{sec:methods-study-design}
\label{sec:methods-rmst-estimand}

SurvFM adapts tabular regression models to static right-censored prediction by constructing restricted mean survival time (RMST) targets from training outcomes. The five configurations were SurvFM-TabPFN, SurvFM-TabICL, SurvFM-TabDPT, SurvFM-TabH2O and SurvFM-MITRA; SurvFM-TabPFN denotes TabPFN-v3 unless stated otherwise. We evaluated the framework in simulations with known conditional RMST, a 55-dataset SurvSet benchmark, a nested training-allocation analysis, held-out prognostic stratification and bidirectional transport between two public myelodysplastic syndrome (MDS) cohorts.

For observation \(i\), let \(X_i\) denote baseline covariates, \(T_i\) event time and \(C_i\) censoring time. The observed data are
\begin{equation}
\label{eq:observed-data}
O_i=(X_i,Y_i,\Delta_i),\qquad
Y_i=\min(T_i,C_i),\qquad
\Delta_i=\mathbb{I}(T_i\le C_i).
\end{equation}
At restriction horizon \(\tau\), the prediction estimand is conditional RMST \cite{royston2013rmst,uno2014beyondhazardratio},
\begin{equation}
\label{eq:conditional-rmst}
\mu_{\tau}(x)=\mathbb{E}\{\min(T,\tau)\mid X=x\}
=\int_{0}^{\tau}S(t\mid X=x)\,dt.
\end{equation}
This is expected restricted event-free time for a covariate profile, not an individually observed event time. In real-data outer training folds, \(\tau_{50}\), \(\tau_{80}\) and \(\tau_{90}\) were the corresponding empirical quantiles of observed event times. The horizon was shared by all models within each dataset, split and horizon block. The \(\tau_{80}\) horizon was primary, with \(\tau_{90}\) and \(\tau_{50}\) as longer and shorter sensitivities. These benchmark horizons are not clinical recommendations; applications should prespecify a clinically relevant horizon with adequate follow-up.

\subsection{SurvFM target construction and model interface}
\label{sec:methods-pseudo-rmst}
\label{sec:methods-backbone-interface}

Let \(\mathcal D\) be an outer training fold with \(n_{\mathcal D}\) observations. The marginal Kaplan--Meier estimator and its RMST functional are \cite{kaplan1958nonparametric}
\begin{align}
\label{eq:km-estimator}
\widehat S_{\mathcal D}(t)
&=\prod_{u\le t}\left\{1-\frac{d_{\mathcal D}(u)}{r_{\mathcal D}(u)}\right\},\\
\label{eq:km-rmst-functional}
\widehat\theta_{\tau}(\mathcal D)
&=\int_{0}^{\tau}\widehat S_{\mathcal D}(t)\,dt,
\end{align}
where \(d_{\mathcal D}(u)\) and \(r_{\mathcal D}(u)\) are the numbers of events and observations at risk at time \(u\). For \(\mathcal D_{-i}\), the fold with observation \(i\) removed, the jackknife pseudo-RMST target is
\begin{equation}
\label{eq:pseudo-rmst}
Z_i(\tau;\mathcal D)
=n_{\mathcal D}\widehat\theta_{\tau}(\mathcal D)
-(n_{\mathcal D}-1)\widehat\theta_{\tau}(\mathcal D_{-i}).
\end{equation}
This construction provides one quantitative target per training observation \cite{andersen2003pseudoregression,andersen2004rmstpseudo}. The marginal construction requires censoring to be unrelated to event time and to predictors entering the regression, with adequate support through \(\tau\) \cite{andersen2010pseudoobservations,parner2010pseudoobservations,overgaard2017asymptotic,overgaard2019pseudoobservations}. Under these conditions,
\begin{equation}
\label{eq:pseudo-expectation}
\mathbb{E}\{Z_i(\tau;\mathcal D)\mid X_i=x\}
=\mu_{\tau}(x)+o_p(1).
\end{equation}
Conditional independence given predictors alone does not justify a target built from an unconditional Kaplan--Meier curve. When censoring depends on prognostic predictors, a stratified, conditional or inverse-probability-weighted target may instead be required. A bounded covariate-dependent-censoring diagnostic is reported in \textbf{Supplementary Table~8}.

Targets were calculated from outer-training outcomes only. Exact leave-one-out curves were evaluated in memory-bounded batches on a common event-time grid. Finite-sample pseudo-observations and final predictions were separately projected to the natural RMST range,
\begin{equation}
\label{eq:target-clipping}
\widetilde Z_i(\tau)=\Pi_{[0,\tau]}\{Z_i(\tau)\},\qquad
\Pi_{[0,\tau]}(u)=\min\{\tau,\max(0,u)\}.
\end{equation}
Target projection was treated as finite-sample stabilization rather than classical pseudo-observation theory; raw-target sensitivity results are in \textbf{Supplementary Table~9}.

For training fold \(\mathcal D_{\mathrm{train}}\), tabular model \(b\) receives
\begin{equation}
\label{eq:survfm-context}
\mathcal C_{\tau}(\mathcal D_{\mathrm{train}})
=\{(X_i,\widetilde Z_i(\tau)):i\in I_{\mathrm{train}}\}.
\end{equation}
Its prediction for held-out observation \(j\) is
\begin{equation}
\label{eq:survfm-prediction}
\widehat m_{\tau,b}(X_j)
=\mathcal A_b\{\mathcal C_{\tau}(\mathcal D_{\mathrm{train}}),X_j\}.
\end{equation}
All five SurvFM models were used as black-box regressors without survival-specific fine-tuning. For discrimination, predicted RMST was sign-reversed,
\begin{equation}
\label{eq:risk-orientation}
\widehat r_{\tau,b}(X_j)=-\widetilde m_{\tau,b}(X_j),
\end{equation}
so higher risk denotes shorter predicted restricted event-free time.

\subsection{Models, preprocessing and prediction procedures}
\label{sec:methods-models}
\label{sec:methods-preprocessing}

The SurvFM models were TabPFN 8.0.8 with its v3 regression checkpoint, TabICL 2.1.1, TabDPT 1.2.0, the hosted TabH2O regression API and the \texttt{Mitra} base model in AutoGluon Tabular 1.5.0 \cite{hollmann2025tabpfn,qu2025tabicl,ma2024tabdpt,pfeiffer2026tabh2o,zhang2025mitra,erickson2020autogluontabular}. Four conventional regressors used the same targets: linear regression, random-forest regression, gradient-boosting regression and XGBoost regression \cite{breiman2001randomforests,friedman2001gradientboosting,chen2016xgboost}. Eight survival comparators comprised ridge Cox proportional hazards, random survival forests, gradient-boosting survival analysis, Weibull and log-normal accelerated failure-time models, DeepSurv, DeepHit and XGBoost-Cox \cite{cox1972regression,ishwaran2008random,hothorn2006survivalensembles,kalbfleisch2002failuretime,katzman2018deepsurv,lee2018deephit}.

Configurations were fixed across datasets rather than tuned separately. TabPFN used one estimator; TabICL and deterministic TabDPT used eight estimators or prediction ensembles. TabH2O imposed a 100-processed-column limit, and MITRA returned the base \texttt{Mitra} prediction. Random forests used 200 trees with minimum leaf size three; boosting comparators used 200 stages; XGBoost regressors used 300 depth-three trees. DeepSurv and DeepHit used hidden layers of 64 and 32 units, a seeded 80/20 inner split and early stopping. CoxPH used a fixed \(L_2\) penalty of \(10^{-4}\) for numerical stability. Failed fits were retained without substitution or imputation. Complete versions, hyperparameters, checkpoints, hardware and service constraints are reported in \textbf{Supplementary Table~3} and the reproducibility files.

Static datasets were loaded with SurvSet 0.2.11 \cite{drysdale2022survset}. Endpoint coding was fixed before splitting. Each public dataset used ten fixed event-stratified 70\%-training and 30\%-test splits with \texttt{random\_state=1000+s}. Within each training fold, numeric variables were median-imputed and standardized, while categorical variables were mode-imputed and one-hot encoded with unseen test levels ignored. The fitted transformation was then applied to the test fold. Test outcomes were not used for preprocessing, horizon selection or target construction.

For survival models returning \(\widehat S_b(t\mid x)\), RMST was
\[
\widehat m_{\tau,b}(x)=\int_0^\tau \widehat S_b(t\mid x)\,dt.
\]
Native curves were evaluated on 200 equally spaced points over \([0,\tau]\), bounded to \([0,1]\), integrated by the trapezoidal rule and projected to \([0,\tau]\). XGBoost-Cox relative risk was combined with an outer-training Breslow baseline \cite{breslow1974covariance}. This conversion placed all models on the same horizon-specific RMST scale.

\subsection{Simulation experiments with known RMST}
\label{sec:methods-controlled-simulation}
\label{sec:methods-paired-censoring}
\label{sec:methods-simulation-summaries}

The main simulation crossed 24 settings, ten replicates and three horizons. Four signal structures were combined with selected values of \(n\in\{50,100,200,500\}\), \(p\in\{10,50,99\}\), censoring and noise (\textbf{Supplementary Table~1}). Independent signal covariates were \(X_1,X_2\sim\mathrm{Uniform}(-1,1)\) and \(X_3,X_4\sim N(0,1)\), with \(p-4\) irrelevant standard-normal covariates. Let \(\eta(X)=s g(X)\), where \(s=1\) under low noise and \(s=0.65\) under high-noise sparse signal, and
\[
\begin{aligned}
g_{\mathrm{linear}}(X)&=X_1-0.8X_2+0.4X_3,\\
g_{\mathrm{nonlinear}}(X)&=1.15\sin(\pi X_1)+X_2^2-0.8X_1X_3,\\
g_{\mathrm{interaction}}(X)&=1.4\mathbb I(X_1>0)+1.6X_1X_2-0.7\mathbb I(X_3>0.5),\\
g_{\mathrm{subgroup}}(X)&=
\begin{cases}-0.6X_1+0.6X_2,&X_4\le0,\\-1.5X_1X_2+X_1^2,&X_4>0.
\end{cases}
\end{aligned}
\]
The first three structures used Weibull proportional hazards with shape \(a=1.5\),
\[
T=\exp\{-\eta(X)/a\}\{-\log(U_T)\}^{1/a},\qquad U_T\sim\mathrm{Uniform}(0,1).
\]
Subgroup settings used \(T=\exp\{\eta(X)+\sigma(X)\varepsilon\}\), where \(\varepsilon\sim N(0,1)\), \(\sigma(X)=0.45\) for \(X_4\le0\) and 0.65 otherwise; high noise multiplied these standard deviations by 1.20. Independent censoring followed \(C\sim\mathrm{Uniform}(0,c_{\max})\), with \(c_{\max}=5.0,2.5,1.4\) for low, medium and high censoring. Data and split seeds for setting \(k\) and replicate \(r\) were \(2026061500+1000k+r\) and \(300000+17k+r\). Each dataset used a 70/30 split, and training-event quantiles defined the horizons. Oracle RMST was obtained from the known survival function on a 1,024-point grid. All 17 models shared data, splits, horizons and oracle values, producing 12,240 attempted rows. Regime labels were overlapping descriptors rather than isolated factorial effects.

The paired censoring intervention held the latent prediction problem fixed. Settings 1, 4, 5, 7, 9, 12, 13, 16, 17, 19, 21 and 24 were selected before model outcomes were examined, with ten replicates per setting. Within a replicate, covariates, latent event times, oracle RMST and train--test membership were generated once; coupled mechanisms then produced exactly 0\%, 20\%, 40\% and 60\% censoring at \(\tau_{80}\). SurvFM-TabPFN and SurvFM-TabICL were trained with known conditional RMST, complete-data restricted time, projected pseudo-RMST, naive restricted observed time or an event-only target. CoxPH and RSF were fitted conventionally at each level. Arms shared preprocessing, held-out observations and seeds. Primary outcomes were oracle RMST error, degradation from uncensored training and gaps to the complete-data and oracle-target references. Setting-cluster percentile intervals used 10,000 resamples retaining censoring levels and replicates (seed 20260826). Of 5,760 planned rows, 5,749 completed; all 11 failures were numerical CoxPH failures. Unsupported discrimination metrics remained missing (\textbf{Supplementary Table~11}).

The target-definition ablation held SurvFM-TabPFN fixed and compared pseudo-RMST with naive \(\min(Y,\tau)\) and event-only targets over the full design. Regime ranks were calculated within each evaluable setting-replicate block among all 17 models. A matched diagnostic compared direct pseudo-target error under independent and bounded covariate-dependent censoring. It crossed four signal structures, two sample-size and dimensionality pairs, two censoring levels, two horizons and two dependence strengths. Matched mechanisms shared covariates, event times, uniforms, splits and latent-time horizons, with positivity checks through the horizon. This assessed finite-sample target behaviour rather than general validity under informative censoring (\textbf{Supplementary Table~8}). A raw-versus-projected SurvFM-TabPFN analysis used identical data, splits, horizons, checkpoint and prediction projection across 720 rows per arm (\textbf{Supplementary Table~9}).

\subsection{Public SurvSet experiments}
\label{sec:methods-survset}
\label{sec:methods-training-allocation}
\label{sec:methods-rmst-stratification}

We screened 59 static SurvSet datasets for one independent row per observation, positive follow-up, binary right censoring and covariates available without outcome-derived information. Longitudinal, repeated-row, multi-state and competing-risk objects were excluded. Fifty-five datasets met the protocol (\textbf{Supplementary Table~2}). The benchmark compared 17 models over ten splits and three horizons, yielding 28,050 attempted rows. TabH2O completed 52 datasets because three exceeded its processed-column limit. Fit failures, service limits and metric-support failures were recorded separately and not imputed.

Primary summaries used \(\tau_{80}\), Uno's C-index from negative predicted RMST and \(\tau\)-normalized inverse-probability-of-censoring-weighted (IPCW) RMST error. Dataset-weighted estimates first averaged evaluable splits within dataset and then weighted datasets equally. Strict-common analyses retained blocks evaluable for every declared model. Intervals used 10,000 whole-dataset bootstrap resamples retaining all ten splits (seed 20260716).

The nested allocation analysis used the 52 datasets with \(N\ge100\), ten fixed outer splits and \(\tau_{80}\). The 30\% test set remained fixed, while nested subsets of the full 70\% training pool contained \(\lfloor0.30N\rfloor\), \(\lfloor0.50N\rfloor\) and all training observations:
\begin{equation}
\label{eq:nested-training-allocations}
\mathcal D_{30}\subset\mathcal D_{50}\subset\mathcal D_{70}.
\end{equation}
Training identifiers were deterministically shuffled within event strata and common prefixes guaranteed nesting. The horizon and evaluation censoring distribution were estimated from \(\mathcal D_{70}\) and fixed across allocations. Preprocessing and pseudo-RMST targets were refit within each subset. This yielded 26,520 attempted rows; intervals used 10,000 whole-dataset resamples retaining splits (seed 20260717).

For prognostic stratification, held-out SurvFM-TabPFN predictions in each of 550 \(\tau_{80}\) blocks were divided deterministically into equal-count low, middle and high tertiles without using outcomes. Group survival was estimated by Kaplan--Meier, observed RMST by exact step-function integration and event probability by \(1-\widehat S_{\mathrm{KM}}(\tau_{80})\). For group \(g\) in block \((d,s)\), grouped discrepancy was
\begin{equation}
\label{eq:grouped-rmst-error}
e_{ds}=\frac{1}{3}\sum_{g=1}^{3}
\left|\frac{\widehat\theta^{\mathrm{KM}}_{dsg}-\overline m_{dsg}}{\tau_{ds}}\right|.
\end{equation}
For each group, we recorded its size, number at risk, zero-survival tails and unsupported positive tails. Sensitivities excluded unsupported blocks or required at least 1, 5, 10 or 20 observations at risk in every tertile. Dataset uncertainty used 10,000 resamples (seed 20260719). These diagnostics assess group-level separation and agreement, not individual calibration.

\subsection{Cross-cohort MDS transport analysis}
\label{sec:methods-mds-transport}

We used two public MDS cohorts with post-sampling all-cause overall survival: the International Working Group for Prognosis in MDS cohort underlying IPSS-M and GSE129828 \cite{bernard2022ipssm,reilly2019mdsmethylation}. From 3,323 IWG records, we retained 2,384 pure-MDS observations with known status and positive follow-up, including 1,177 deaths. GSE129828 contributed 116 of 141 records, including 86 deaths; two otherwise eligible censored records with zero follow-up were excluded. GSE129828 survival was interpreted from biospecimen collection because some observed survival times were shorter than the diagnosis-to-sampling interval. Both cohorts contributed one baseline profile per observation. A fingerprint screen found no high-information matches, although anonymization precluded definitive exclusion of overlap (\textbf{Supplementary Table~10}).

Seven clinical predictors were age at sampling, sex, haemoglobin, platelet count, white-cell count, absolute neutrophil count and bone-marrow blast percentage. The molecular representation added 20 shared mutation indicators. Karyotype was excluded because cohort encodings could not be harmonized without interpretation. All preprocessing was estimated in the source cohort. Continuous variables were source-median-imputed with missingness indicators and standardized using source moments; categorical coding and column order followed a source-derived dictionary that was applied unchanged to the target cohort.

Pseudo-RMST targets and all 17 models were fitted using source outcomes only and applied to the target without refitting, target-outcome-guided tuning or substitution. Both directions, both predictor sets and fixed two- and three-year horizons yielded eight configurations per model and 136 analyses. Source-cohort Kaplan--Meier estimates supplied targets and primary censoring weights; target-cohort weights were an evaluation sensitivity. Primary metrics were Uno's C-index and \(\tau\)-normalized IPCW RMST RMSE. Target-observation intervals used 10,000 common-row bootstrap resamples with the source-fitted model held fixed (seed 2026081201), and therefore exclude source-refitting uncertainty. Directional comparisons cannot isolate source sample size because cohort identity also changes.

\subsection{Evaluation, statistical analysis and reproducibility}
\label{sec:methods-evaluation-metrics}
\label{sec:methods-ranking}
\label{sec:methods-reproducibility}

Simulation error was oracle conditional-RMST RMSE,
\begin{equation}
\label{eq:rmse-oracle}
\mathrm{RMSE}_{\mathrm{oracle}}(\tau)
=\left[|J|^{-1}\sum_{j\in J}
\{\widetilde m_{\tau}(X_j)-\mu_{\tau}(X_j)\}^{2}\right]^{1/2},
\end{equation}
where \(J\) is held out and \(\mu_{\tau}\) is known. Realized restricted-time RMSE was secondary and retained in Source Data.

Uno's C-index \cite{uno2011cstatistics} used risk \(\widehat r_{\tau}=-\widetilde m_{\tau}\):
\begin{equation}
\label{eq:uno-c}
\widehat C_{\mathrm{Uno}}\{\widehat r_{\tau}\}
=\frac{\sum_{i,j\in J}\mathbb{I}(Y_i<Y_j,\Delta_i=1)
\mathbb{I}\{\widehat r_{\tau}(X_i)>\widehat r_{\tau}(X_j)\}
\widehat G(Y_i-)^{-2}}
{\sum_{i,j\in J}\mathbb{I}(Y_i<Y_j,\Delta_i=1)\widehat G(Y_i-)^{-2}},
\end{equation}
where \(\widehat G\) was estimated from outer-training outcomes. We used \texttt{concordance\_index\_ipcw} without optional truncation; \(\tau\) identifies the RMST prediction horizon rather than a truncated concordance estimator.

Public-data RMST error adapted IPCW prediction-error principles \cite{graf1999prognostic,gerds2006brierscore}. Define
\begin{align}
\label{eq:ipcw-target-weight}
A_j(\tau)&=\mathbb I(\Delta_j=1,Y_j\le\tau),\\
H_j(\tau)&=\mathbb I(Y_j\ge\tau)\{1-A_j(\tau)\},\\
R_j(\tau)&=\min(Y_j,\tau),\\
w_j(\tau)&=\frac{A_j(\tau)}{\widehat G(Y_j-)}+\frac{H_j(\tau)}{\widehat G(\tau)}.
\end{align}
The normalized metric was
\begin{equation}
\label{eq:ipcw-rmst-rmse}
\mathrm{nRMSE}_{\mathrm{IPCW}}(\tau)
=\left[\frac{1}{|J|}\sum_{j\in J}w_j(\tau)
\left\{\frac{\widetilde m_{\tau}(X_j)-R_j(\tau)}{\tau}\right\}^{2}\right]^{1/2}.
\end{equation}
Rows censored before \(\tau\) contributed zero to the numerator but remained in the full test denominator. We used \(\widehat G(Y_j-)\) for events and \(\widehat G(\tau)\) for horizon-reaching observations, without weight truncation, flooring or self-normalization. A block was non-evaluable when a required \(\widehat G\le10^{-6}\), a weighted comparable-pair denominator was zero, event times exceeded censoring support or predictions were nonfinite. Uno-C similarly required a finite estimate, nonzero weighted comparable-pair denominator and event times within training-fold support. Secondary metrics were Harrell's C-index and cumulative dynamic AUC.

Within-block ranks used competition ranking, with lower RMST error and higher discrimination favorable. Available-evaluable and strict-common analyses were generated separately. Bootstrap schemes followed the statistical unit and used 10,000 percentile resamples. Simulation intervals resampled replicates within settings with common model indices; public analyses resampled whole datasets while retaining splits. Selected exploratory contrasts used paired block differences, fixed-seed bootstrap intervals and Wilcoxon signed-rank tests with Benjamini--Hochberg adjustment \cite{benjamini1995fdr}. Fit failures, hosted-input limits, nonfinite predictions and support failures remained distinct and were not imputed.

Observation-level files retained stable identifiers, outcomes, predictions and support fields, enabling metric reconstruction. Exact pseudo-RMST tests covered censoring-free, all-censored, tied-time and horizon-edge cases. Source-data files were checked for duplicate keys, metadata agreement, deterministic generation and file integrity. Runtime was the median fit-and-prediction time among successful rows; preprocessing and target construction were cached separately. Hosted request latency and local GPU or CPU timings are descriptive rather than controlled speed comparisons. The reproducibility package provides complete seeds, software environments, evaluability rules and links between each figure panel and its source data.

\section{Data availability}

This study used public SurvSet datasets, available through the SurvSet collection and the original sources listed in \textbf{Supplementary Table~2}, and simulated datasets generated by the simulation code. The MDS transport analysis used the IWG cohort available from cBioPortal (study \texttt{mds\_iwg\_2022}; \url{https://www.cbioportal.org/study/summary?id=mds_iwg_2022}) and GSE129828 available from the Gene Expression Omnibus (\url{https://www.ncbi.nlm.nih.gov/geo/query/acc.cgi?acc=GSE129828}) \cite{bernard2022ipssm,reilly2019mdsmethylation}. Numerical source data underlying the figures and tables, benchmark summaries, eligibility and denominator records, harmonization dictionaries, transport results, and split and simulation seed records are provided with the submission as Supplementary Data. Simulated data can be regenerated using the supplied code. No protected health information or newly collected human-participant data are included. A versioned source-data and reproducibility release will be deposited in a public DOI-minting repository no later than publication.

\section{Ethics statement}

The study analysed public, deidentified secondary data and simulated data. No participants were recruited and no new participant data were collected at The University of Texas MD Anderson Cancer Center; institutional review board approval and informed consent were therefore not required for this analysis.

\section{Code availability}

The public SurvFM implementation, including the lightweight \texttt{survfm} package, installation guidance, examples, unit tests and API documentation, is available under the MIT License at \url{https://github.com/yuelyu21/SurvFM}. Version 0.2.0 is archived on Zenodo at \url{https://doi.org/10.5281/zenodo.21735311}; the all-versions DOI is \url{https://doi.org/10.5281/zenodo.21735310}. The rendered documentation is available at \url{https://yuelyu21.github.io/SurvFM/}. The custom benchmark and MDS transport code, harmonization dictionary, software-environment records, figure scripts and quality-control reports are provided to editors and reviewers with the submission.

\bibliographystyle{unsrt}
\bibliography{survfm_references}

\section{Acknowledgements}

We thank members of the study team and collaborators for discussions on survival prediction, benchmark design and interpretation. Y.L. and Z.L. were partially supported by the National Institute of General Medical Sciences (R35GM159819), the National Human Genome Research Institute (R01HG014774) and the National Cancer Institute (R21CA312466). Z.L. was additionally supported by the National Cancer Institute (U24CA274212). X.H. was partially supported by the National Cancer Institute (R01CA272806) and the Dr. Mien-Chie Hung and Mrs. Kinglan Hung Endowed Professorship.

\section{Author contributions}

Y.L. and Z.L. conceived the study. Y.L. developed the methodology and software, conducted all analyses, generated the figures and reproducibility materials, and drafted the manuscript. X.H. provided methodological guidance. Z.L., X.H. and S.H.L. critically reviewed and revised the manuscript. All authors approved the submitted version.

\section{Competing interests}

The authors declare no competing interests.

\section{Additional information}

Supplementary information accompanies this paper. Correspondence and requests for materials should be addressed to X.H. or Z.L.

\clearpage
\section*{Figure legends}

\clearpage
\section*{Extended Data figures}
\setcounter{figure}{0}
\renewcommand{\figurename}{Extended Data Fig.}
\renewcommand{\theHfigure}{ED.\arabic{figure}}

\begin{figure}[p]
\centering
\includegraphics[width=\linewidth,height=0.70\textheight,keepaspectratio]{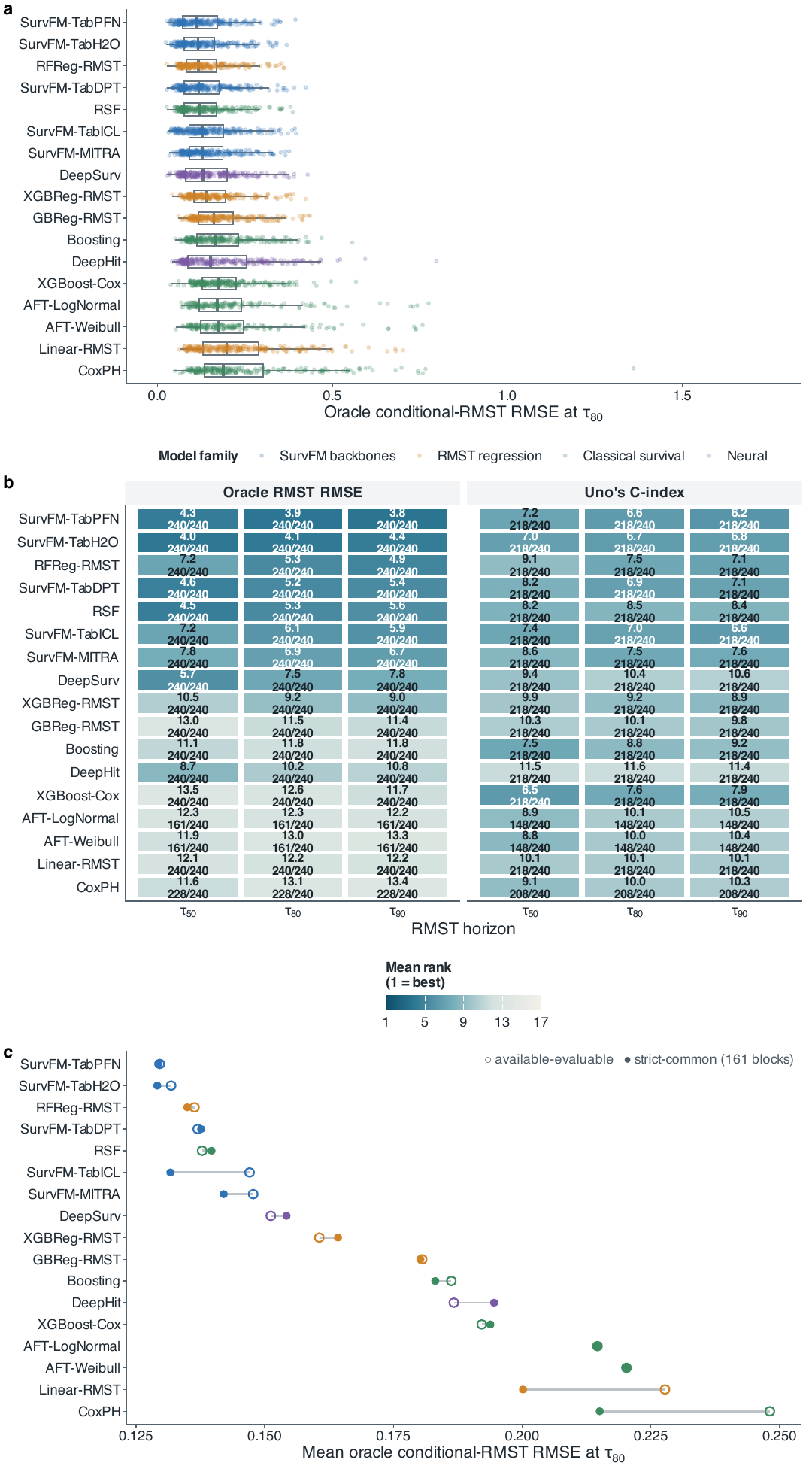}
\caption{\textbf{Full 17-model simulation performance and denominator sensitivity.} The analyses use the fixed 24-setting, ten-replicate simulation design. \textbf{a,} Full distribution of oracle conditional-RMST RMSE at $\tau_{80}$. Points are evaluable setting and replicate blocks; boxes show the median and interquartile range. All values are retained on the displayed 0--1.75 scale. \textbf{b,} Mean within-block ranks for oracle conditional-RMST RMSE and Uno's C-index across $\tau_{50}$, $\tau_{80}$ and $\tau_{90}$. Ranks were calculated among models evaluable in each block. Each cell reports the mean rank and evaluable/attempted block count; rank 1 is best. \textbf{c,} Sensitivity of mean $\tau_{80}$ oracle conditional-RMST RMSE to the analysis denominator. Open markers show each model's available-evaluable mean and filled markers show the mean over the 161 setting and replicate blocks where all 17 models were evaluable. The strict-common set removes 79 of 240 blocks, concentrated at $n=50$ and $n=100$, and conditions the analysis on accelerated failure-time-model convergence. Missing fits or metrics were not imputed.}
\label{edfig:simulation-complete}
\end{figure}

\begin{figure}[p]
\centering
\includegraphics[width=\linewidth,height=0.78\textheight,keepaspectratio]{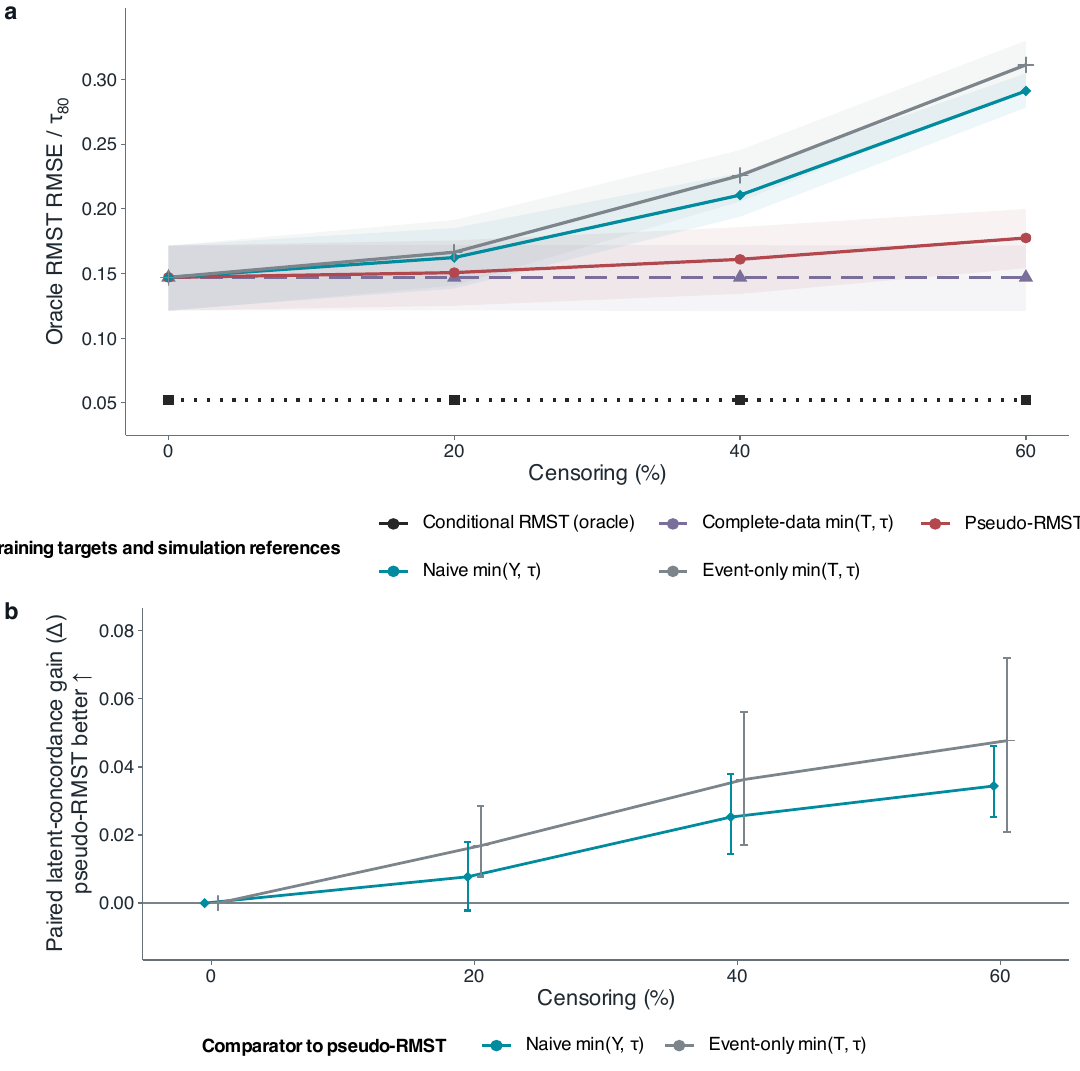}
\caption{\textbf{Complete-data and conditional-RMST references for the censoring intervention.} Covariates, latent event times, train--test splits and horizons were coupled across exactly 0\%, 20\%, 40\% and 60\% censoring. \textbf{a,} Mean oracle RMST RMSE divided by $\tau_{80}$ for SurvFM-TabPFN trained with projected pseudo-RMST, naive $\min(Y,\tau)$, event-only $\min(T,\tau)$, complete-data $\min(T,\tau)$ and conditional-RMST reference targets. Ribbons are 95\% setting-cluster bootstrap intervals. The complete-data arm removes censoring while retaining realized event-time variability; the conditional-RMST arm supplies the true regression target. \textbf{b,} Paired difference in latent-time concordance between projected pseudo-RMST and the two observed-outcome alternatives. Positive values favour projected pseudo-RMST. Points are mean paired differences and bars are 95\% setting-cluster bootstrap intervals.}
\label{edfig:censoring-references}
\end{figure}

\end{document}